\documentclass{article}
\usepackage{blindtext}
\usepackage[T1]{fontenc}
\usepackage[utf8]{inputenc}
\usepackage{amsmath, amssymb}
\usepackage[]{algorithmic}
\usepackage{graphicx}
\usepackage{authblk}

\title{	Projection matrices and related viewing frustums: new ways to create
		and apply}

\author[]{Nikita Glushkov\thanks{nike.glushkov@gmail.com} }
\author[]{Tyuleneva Emiliya\thanks{strongamil1998@gmail.com}}
\affil{Graphical Department\\ Huawei Russian Research Institute\\ Saints-Petersburg Russian Federation}
\renewcommand\Authands{ and }
\begin{document}
			
\maketitle
\begin{abstract}
In computer graphics, the field of view of a camera is represented by a viewing frustum and a corresponding projection matrix, the properties of which, in the absence of restrictions on
rectangular shape of the near plane and its parallelism to the far plane are currently not fully explored and structured.\\
This study aims to consider the properties of arbitrary affine frustums, as well as various techniques for their transformation for practical use in devices with limited resources.
Additionally, this article explores the methods of working with the visible volume as an arbitrary frustum that is not associated with the projection matrix.\\
To study the properties of affine frustums, the dependencies between its planes and formulas for obtaining key points from the inverse projection matrix were derived. Methods of constructing frustum by key points and given planes were also considered. Moreover, frustum transformation formulas were obtained to simulate the effects of reflection, refraction and cropping in devices with limited resources. In conclusion, a method was proposed for applying an arbitrary frustum, which does not have a corresponding projection matrix, to limit the visible volume and then transform the points into NDC space.
\end{abstract} \hspace{10pt}
	\section{Introduction}
	There are several ways to represent the visible volume in computer graphics, and one of them is to store it as a viewing frustum\cite{first}. From the mathematical point of view, it can be represented by three pairs of planes. If for any two taken pairs of planes it is true that they intersect at one point (the point at infinity is also taken into account here), then such a frustum can be represented by an affine 4x4 matrix. We will call such frustums affine, otherwise - non-affine. Affine frustums are considered in more detail in Chapter~\ref{section:1}, and the non-affine ones in Chapter~\ref{section:nonAffine}. 
	\\It is also worth noting that in this article we use matrices with the order of multiplication: vector~*~matrix. If we need the inverse version of multiplication, then the matrix must be transposed. To conduct perspective projection when converting points to NDC space, it is considered that along the OX and OY axes in the clip space, they belong to the range [-1, 1], and along the OZ axis to the [0, 1]. Thus, all the formulas below will be correct for the specified transformation, however, following the analogy, they can be reproduced for other accepted coordinate systems.
	\\In addition, it is necessary also to mention the efficiency of storing affine frustum as a pair of matrices: projection matrix $M$ and its inverse $M'$, which makes it possible to quickly perform matching of points between projective space and three-dimensional space.
	\\The main goal of this work is to consider the properties of arbitrary affine frustums, as well as algorithms for their construction and transformation with the aim of further practical application in the implementation of such
	effects as reflection, refraction, truncation of the visible volume for application in devices with limited resources. In addition, this article describes the idea
	of using non-affine frustums in computer graphics, which describe the visible volume of a scene, in order to 		 
	eliminate the problem of dependencies between planes in traditional affine frustums.
\section{Related Work}
In a related paper \cite{second}, the authors investigated the construction of projection matrices both for the case of the far plane going to infinity, and for the case of cutting off frustum by a new near plane, in which the far plane is maximally distanced. However, in this work, only standard OpenGL matrices were considered, and no generalization to an arbitrary case was carried out. Gribb and Hartmann in their work \cite{third} described the relationship between the projection matrix and the 6 planes of the pyramid of visibility, which was a fundamental foundation for further investigation of the properties of projection matrices.
\section{Relationship between affine frustum and projection matrix}
\label{section:1}
\subsection{Retrieving clue planes from projection matrix}\label{chapter:2.1}
As mentioned earlier, there is a dependency between the 6 planes of the visibility pyramid: left, right, top, bottom, near, far, denoted as $\vec{L}, \vec{R}, \vec{T}, \vec{B}, \vec{N} , \vec{F}$ respectively. It is possible for these planes to be received from the corresponding projection matrix $M(\vec{m_0}, \vec{m_1}, \vec{m_2}, \vec{m_3})$ as follows. Based on \cite{third}, we can obtain 6 frustum planes from following equation:

\begin{equation}
	\begin{cases}
	\vec{m_0} - \vec{m_3} = K_r \cdot \vec{R},\\
	\vec{m_0} + \vec{m_3} = K_l \cdot \vec{L},\\
	\vec{m_1} - \vec{m_3} = K_t \cdot \vec{T},\\
	\vec{m_1} + \vec{m_3} = K_b \cdot \vec{B},\\
	\vec{m_2} - \vec{m_3} = K_f \cdot \vec{F},\\
	\vec{m_2}  = K_n * \vec{N}.
	\end{cases}
	\label{sys:1}
\end{equation}

\begin{figure}[h]
	\centering
	\begin{minipage}{0.4\textwidth}
		\includegraphics[width=\linewidth]{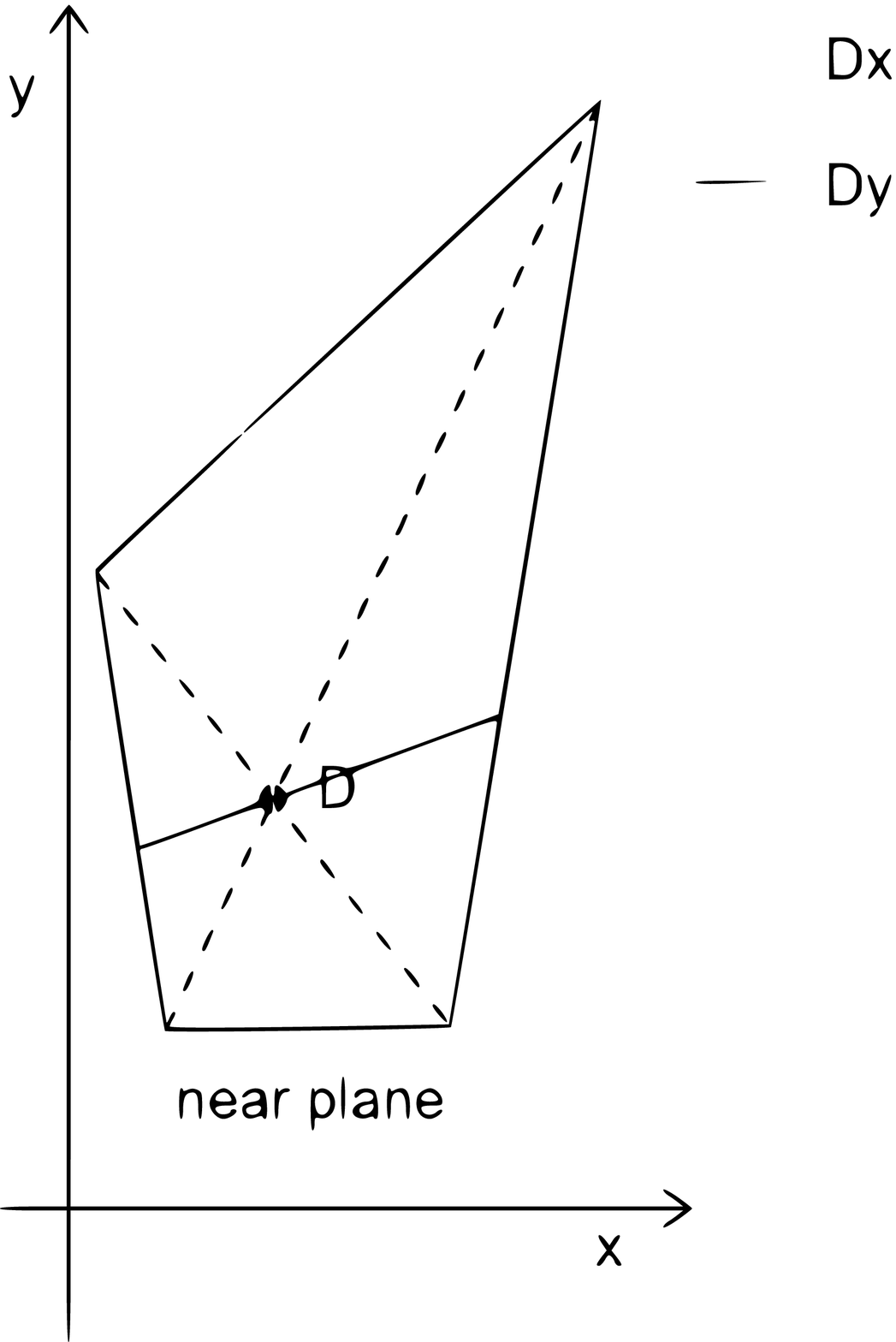}
		\caption{
			\label{planes}
			Planes of coordinate-wise division of the affine frustum in the world coordinate system}
	\end{minipage}\hfill
	\begin{minipage}{0.55\textwidth}
		\includegraphics[width=\linewidth]{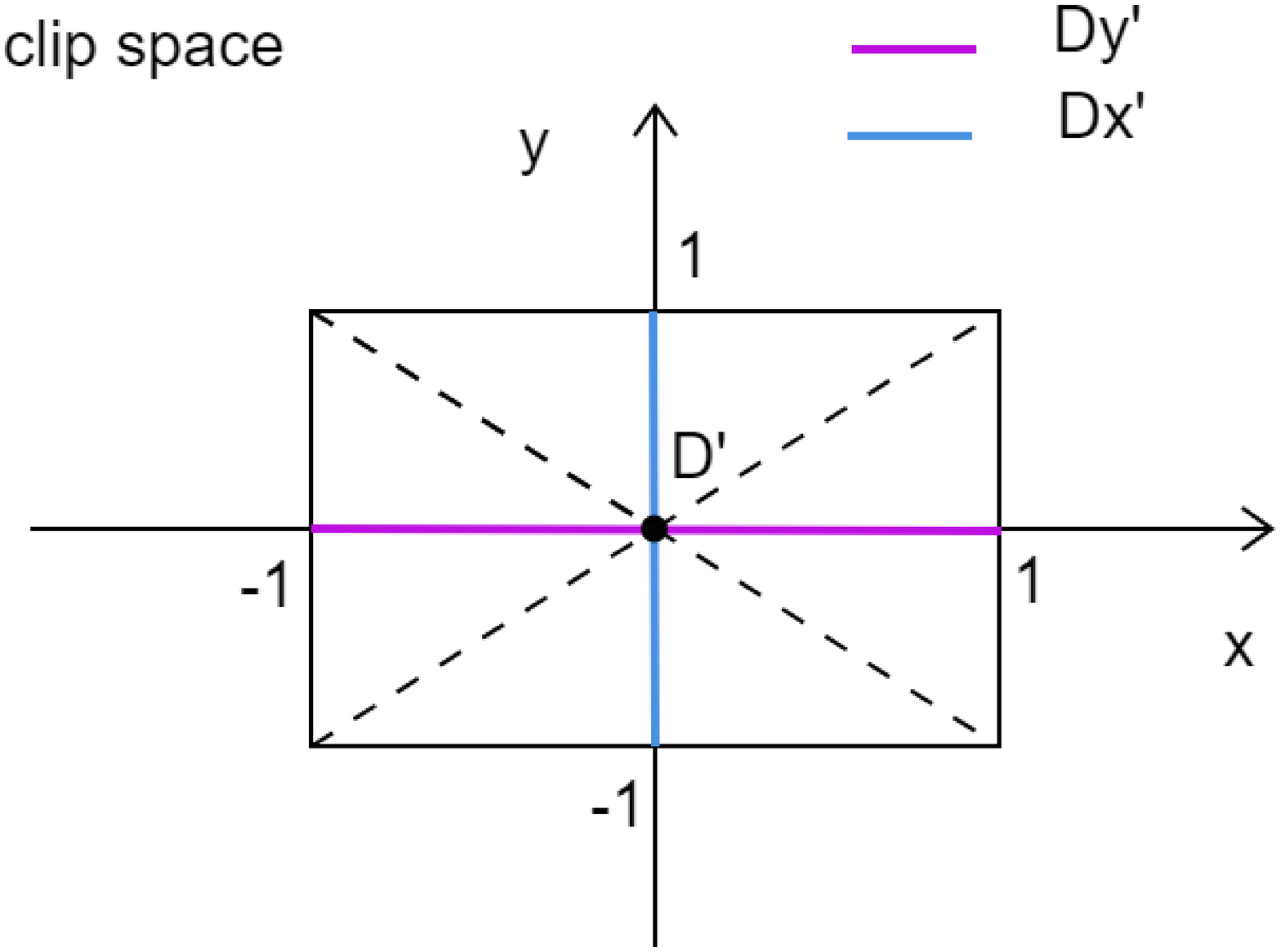}
		\caption{Planes of coordinate-wise division of the affine frustum in the clip space}\label{clip}
	\end{minipage}
\end{figure}

It is assumed that  $\vec{L}, \vec{R}, \vec{T}, \vec{B}, \vec{N} , \vec{F}$ are fixed by us, and taken with some unknown coefficients $K_i$. As you can see, the last equation in \eqref{sys:1} differs from the others due to the fact that the range along the OZ axis ([0,~1]) in clip space is different from the rest of the axes([-1,~1]). By transforming the system above, the following expressions can be written:
\begin{equation}
\begin{cases}
	2\vec{m_0} = K_r \cdot \vec{R} + K_l \cdot \vec{L},\\
	2\vec{m_1} = K_t \cdot \vec{T} + K_b \cdot \vec{B},\\
	2\vec{m_3} = K_l \cdot \vec{L} - K_r \cdot \vec{R},\\
	2\vec{m_3} = K_b \cdot \vec{B} - K_t \cdot \vec{T},\\	
	\vec{m_2} - \vec{m_3} = K_f \cdot \vec{F},\\
	\vec{m_2}  = K_n \cdot \vec{N}.
	\label{eq:2}
\end{cases}
\end{equation}
Thus, for an affine frustum, there are the following restrictions, which it must satisfy:
\begin{equation}
	\begin{cases}
		K_n \cdot \vec{N} - (0.5 K_l \cdot \vec{L} + 0.5 K_r \cdot \vec{R}) = K_f \cdot \vec{F},\\
		K_l \cdot \vec{L} - K_r \cdot \vec{R} = K_b \cdot \vec{B} - K_t \cdot \vec{T}.\label{eq:3}
	\end{cases}
\end{equation}
The second equation in \eqref{eq:3} will be discussed and solved later in Chapter \ref{ch:4}.
\\In a similar way, the equations of the planes $\overrightarrow{D_x}, \overrightarrow{D_y}$ (Figure~\ref{planes}), dividing the affine frustum in clip space along each of the axes in half (Figure~\ref{clip}) and passing through its central point $D$ (intersection point of its diagonals), in the world coordinate system can be obtained using the following formulas:
\begin{equation}
	\begin{cases}
		\vec{m_0} = \overrightarrow{Dx},\\
		\vec{m_1} = \overrightarrow{Dy},\\
		\vec{m_2} - 0.5\vec{m_3} = \overrightarrow{Dz}.\\
	\end{cases}
\end{equation}

\subsection{Retrieving key points from inverted projection matrix
}
\begin{figure}
			\centering
			\begin{minipage}{0.4\textwidth}
			\includegraphics[width=\linewidth]{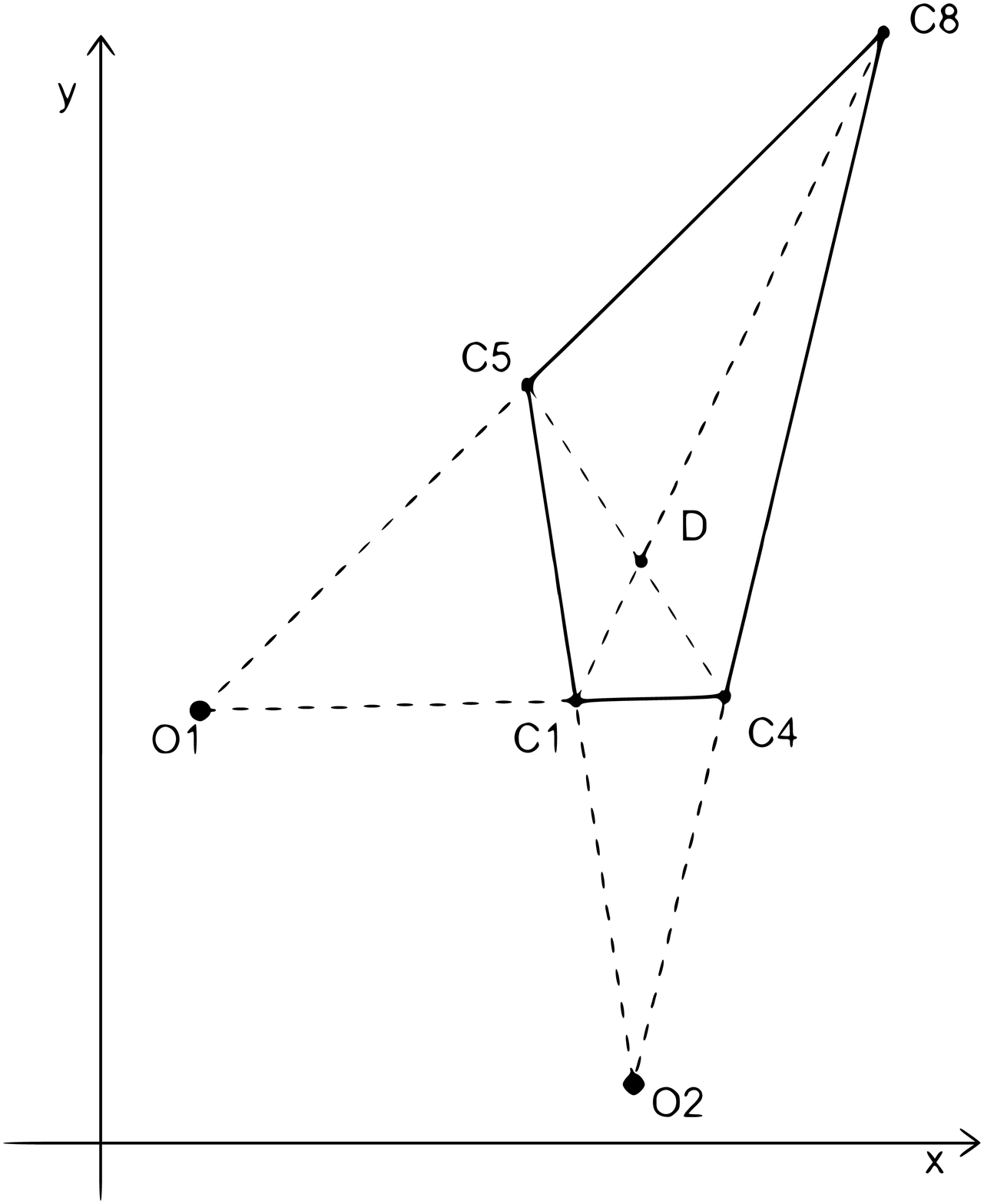}
			\caption{Vanishing points $O_1$, $O_2$ and central point D}\label{qlue_dots}
			\end{minipage}\hfill
				\begin{minipage}{0.5\textwidth}
			\includegraphics[width=\linewidth]{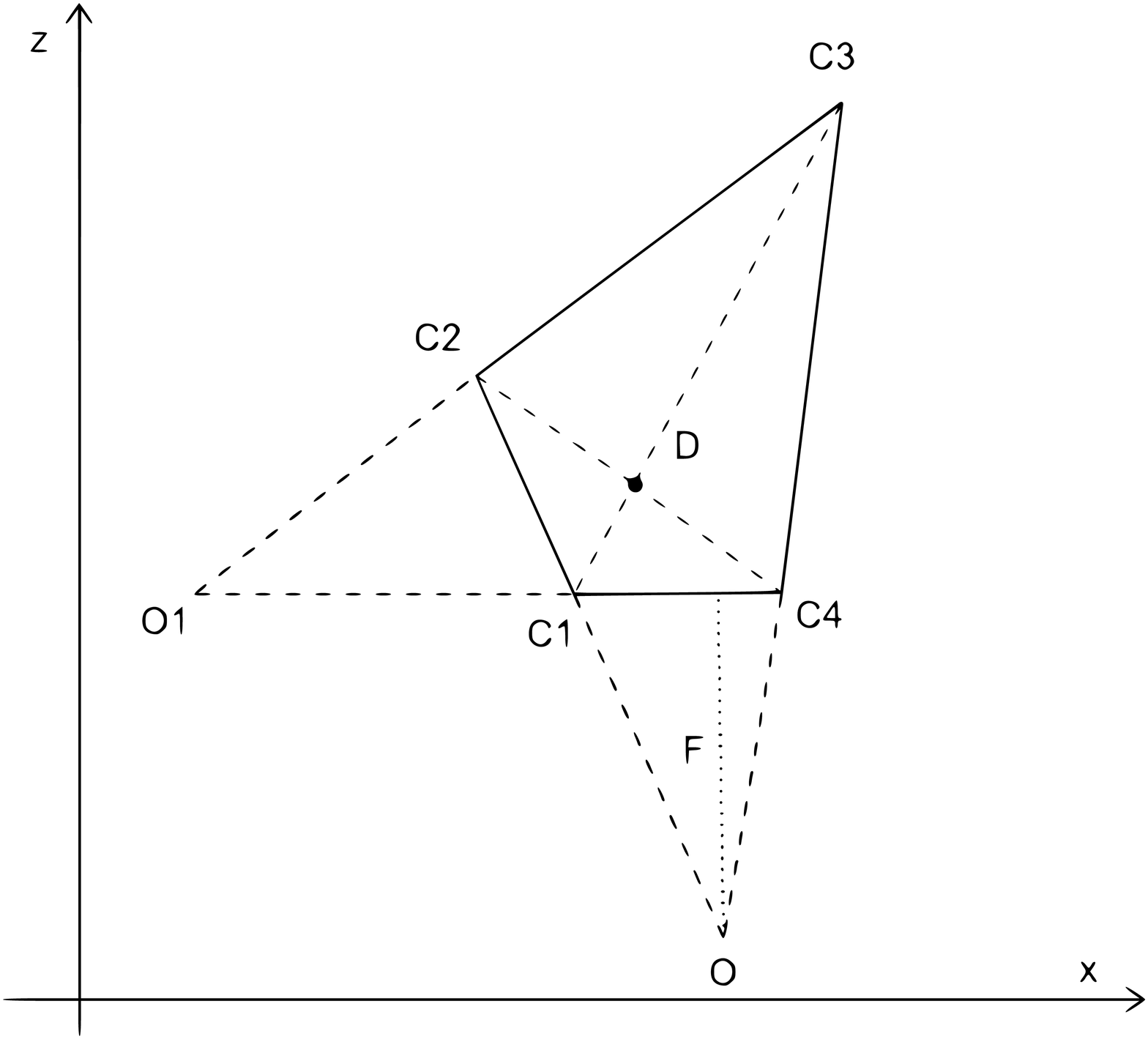}
			\caption{Vanishing points $O_1$, $O$ (camera position) and central point D}\label{key2}
		\end{minipage}
		\end{figure}

\begin{figure}[ht]
		\begin{center}
			\scalebox{0.3}{
				\includegraphics{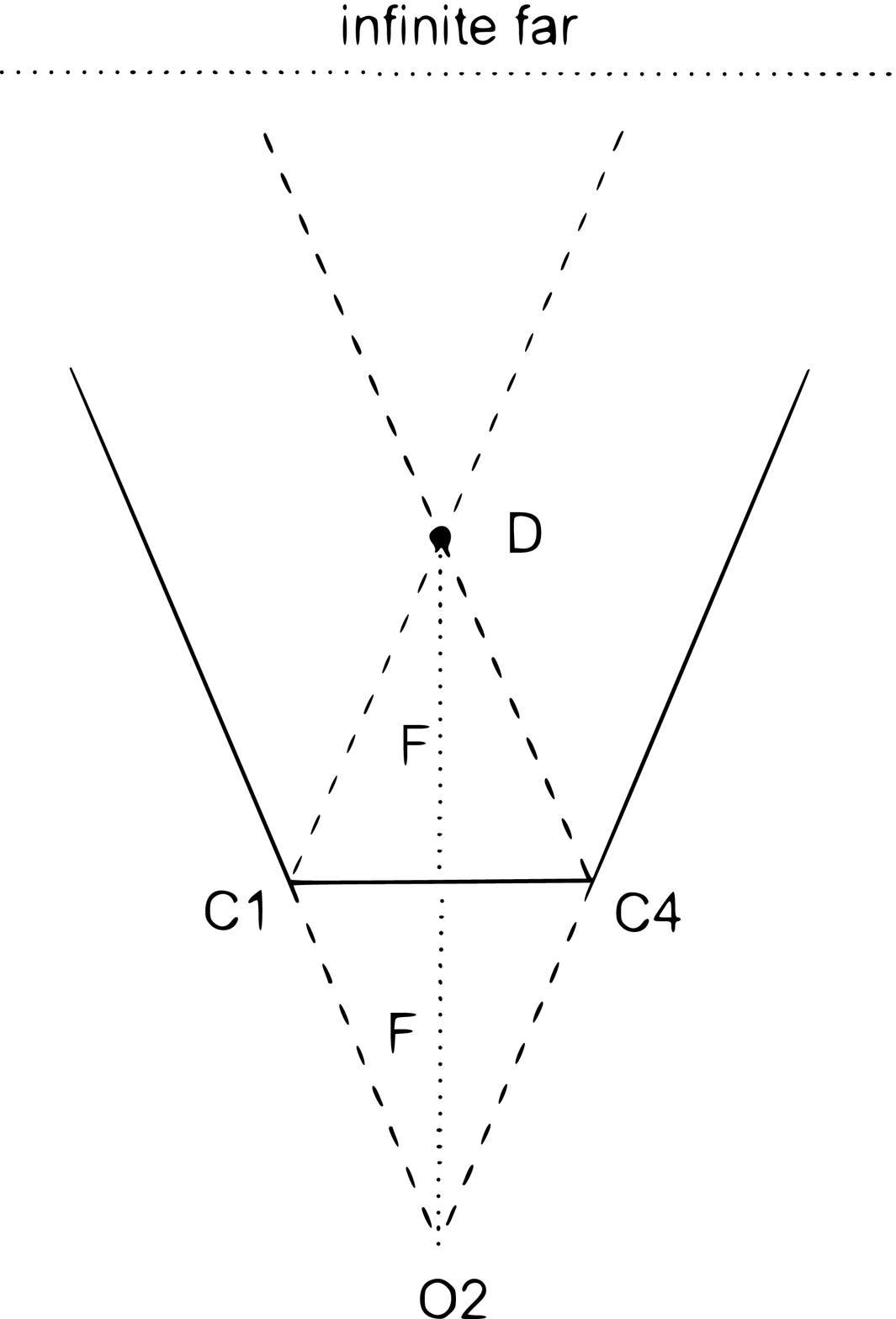}
			}		
			\caption{		\label{inf}			
				Case of infinite far plane}			
		\end{center}
	\end{figure}

In some cases it can be useful to quickly obtain key points (Figure~\ref{qlue_dots},~\ref{inf}) of the affine frustum: its corner points $C_1...C_8$, central point $D$  (intersection point of its diagonals) and 3 vanishing points $O, O_1, O_2$($O$ - camera position) in world coordinates. 
	
This can be done by multiplying the corresponding homogeneous coordinates of the desired point in NDC by the inverse projection matrix $M'\begin{pmatrix} \vec{m_0}'\\ \vec{m_1}'\\ \vec{m_2}'\\ \vec{m_3}'\end{pmatrix}$, what has been done in \eqref{eq:7}, \eqref{eq:8} and after converting it back to 3D space. 
Since we use homogeneous coordinates, the desired point can be even moved to infinity, such as the position of the camera in an orthogonal projection.
\\In order to obtain homogeneous coordinates of vanishing points, consider the corresponding lines $l, l_1, l_2$ in the clip space, as a set of parameterized points\eqref{eq:5}. When moving along a given line to infinity, the parameter $t\to\infty$, and in the limit we obtain the desired point in homogeneous coordinates \eqref{eq:6}.
\begin{equation}
		\begin{cases}
				l = \{(0, 0, t, 1)\} \sim \{(0, 0, 1, \frac{1}{t})\} \\
				l_1 = \{(0, t, 0.5, 1)\} \sim \{(0, 1, \frac{0.5}{t}, \frac{1}{t})\}\\
				l_2 = \{(t, 0, 0.5, 1)\} \sim \{(1, 0, \frac{0.5}{t}, \frac{1}{t})\} \label{eq:5}
		\end{cases} 
	\end{equation}

	\begin{equation}
	\begin{cases}
		O' = (0, 0, 1, 0) \\O_1' = (0, 1, 0, 0)\\O_2' = (1, 0, 0, 0) \label{eq:6}
	\end{cases}
	\end{equation}
\begin{equation}
\begin{cases}
		O = k*\vec{m_2}', k \in Z\\O_1 = k_1*\vec{m_1}', k_1 \in Z\\O_2 = k_2*\vec{m_0}', k_2 \in Z \label{eq:7}
	\end{cases}
\end{equation}
Homogeneous coordinates of corner $C_i'$ and center points $D'$ are calculated quite simply.
\begin{equation}
	\begin{cases}
	 C_1' = (1, 1, 0, 1) \\ C_2' = (-1, 1, 0, 1) \\ C_3' = (-1, -1, 0, 1) \\ C_4' =(1, -1, 0 , 1)\\
		C_5' = (1, 1, 1, 1) \\ C_6' = (-1, 1, 1, 1) \\ C_7' = (-1, -1, 1, 1) \\ C_8' =(1, -1, 1, 1)\\D' =(0, 0, 0.5, 1)
	\end{cases} \implies
	\begin{cases}
 C_1 = \vec{m_0}' + \vec{m_1}' + \vec{m_3}' \\ C_2 = -\vec{m_0}' + \vec{m_1}' + \vec{m_3}' \\ C_3 = -\vec{m_0}' - \vec{m_1}' + \vec{m_3}' \\ C_4 = \vec{m_0}' - \vec{m_1}' + \vec{m_3}'\\
	C_5 = \vec{m_0}' + \vec{m_1}' + \vec{m_2}' + \vec{m_3}' \\ C_6 = -\vec{m_0}' + \vec{m_1}' + \vec{m_2}' + \vec{m_3}' \\ C_7 = -\vec{m_0}' - \vec{m_1}' + \vec{m_2}' + \vec{m_3}' \\ C_8 = \vec{m_0}' - \vec{m_1}' + \vec{m_2}' + \vec{m_3}'\\D = 0.5\vec{m_2}' + \vec{m_3}' \label{eq:8}
\end{cases}
\end{equation}
\\Due to the fact that point D in a non-degenerate affine frustum cannot be moved away from the near plane further than by the focus value (the distance from the camera to the near plane), the problem of non-uniform distribution of z values in perspective space arises\cite{fifth}.
\section{Affine frustum construction from planes}\label{ch:4}
Consider the inverse problem of constructing the projection matrix $M$ of an affine frustum from given side ($\vec{L}, \vec{R}, \vec{T}, \vec{B}$) and near $\vec{N}$ planes. It can be done by using \eqref{eq:2}, however it is necessary to calculate the unknown coefficients $K_i$.

Due to the fact that the projection matrix can be determined up to multiplication by a coefficient other than zero, we can fix one of the unknown coefficients: $K_l = 1$. Then second equation in \eqref{eq:3} can be solved if the given lateral planes really intersect at one point - camera position $O$:
\begin{equation}
\begin{pmatrix} \vec{L}\\ \vec{R}\\ \vec{T}\\ \vec{B}\end{pmatrix}\begin{pmatrix}\vec{O} \\1 \end{pmatrix}= 0 \label{sys:9}
\end{equation}

Indeed if it is so, then \eqref{sys:9} has a nonzero solution and the determinant of its matrix equals 0. By the corollary of the basic minor theorem vectors $ \vec{L}, \vec{R}, \vec{T}\, \vec{B}$ are linearly dependent and second equation in \eqref{eq:3} is solvable. 
So the coefficients, introduced in \ref{chapter:2.1} can be calculated as follows:
\begin{equation}
	\begin{pmatrix} Kr\\ Kb\\ Kt \end{pmatrix} = \begin{pmatrix}
	R_x & B_x & -T_x \\ 
	R_y & B_y & -T_y \\
	R_z & B_z & -T_z \end{pmatrix}^{-1} * \begin{pmatrix} L_x\\ L_y \\ L_z \end{pmatrix}
\end{equation}

After retrieving $K_r, K_l, K_b, K_t$, it remains necessary to calculate the coefficient $K_n$ to calculate the third column of the projection matrix. It must be remembered that this coefficient affects the location of the far plane of the affine frustum due to the relationship between all its planes, therefore it must be chosen from the conditions imposed on it.
\subsection{Point belonging to a far plane}
\begin{figure}[ht]
	\begin{center}
		\scalebox{0.3}{
			\includegraphics{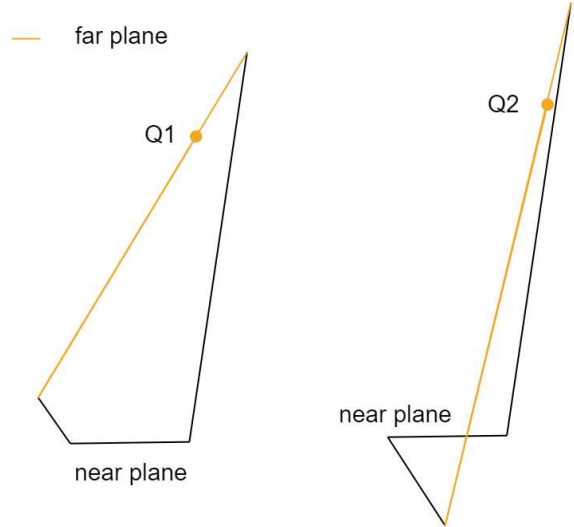}
		}		
		\caption{
			\label{deg}			
			Deflection of the far plane when moving the point $Q_1$ passing through it to a new point $Q_2$, which is farther from the near plane than the starting point $Q_1$}
		
		\end {center}
		\end {figure}

One of the possible restrictions is the belonging of given point Q  the far plane, then $K_n$ can be obtained from the equations:
\begin{equation}
	(K_n * \overrightarrow{N} - \overrightarrow{m_3}) = \vec{F}
	\label{eq:11}
\end{equation}
\begin{equation}
(K_n * \overrightarrow{N} - \overrightarrow{m_3}) \cdot \overrightarrow{Q} = 0
\end{equation}
\begin{equation}
	K_n = \frac{\overrightarrow{m_3} \cdot \overrightarrow{Q}}{\overrightarrow{N} \cdot \overrightarrow{Q}}
\end{equation}
It should be noted that this method is not universal for arbitrary frustum with non-rectangular near plane: when the point Q moves too far away from the near plane, there is a strong deflection of the far plane and degeneration of the frustum (Figure~\ref{deg}).
	\subsection{Increase in volume between the near and far planes}
	The explicit solution to the volume increase problem between the near and far planes is generally nontrivial and requires too many calculations, since it becomes necessary to apply numerical methods.
	To find a non-optimal, but feasible solution, it is proposed to use the following method:
	\begin{enumerate}
		\item First, we need to calculate the coefficient $K_n$ for the case when the far plane becomes
		perpendicular to the near \eqref{eq:13}.
		\item Next, it is necessary to iteratively increase the obtained coefficient $K_n$ in order to smoothly change the vector of the far plane plane $\vec{F}$ in Equation~\eqref{eq:11} in the direction of approaching the vector of the near plane N, which makes it possible to return the frustum to a state where it has not yet become degenerate.
	\end{enumerate}	
\begin{equation}
	K_n = \frac{(\overrightarrow{m_{3xyz}}, \overrightarrow{N_{xyz}})}{|\overrightarrow{N_{xyz}}|^2} \label{eq:13}
	\end{equation}
The criterion for stopping the modification of the far plane $\vec{F}$ is the location of the four near corner points $C_1, C_2, C_3, C_4$ on one side of the far plane.
\begin{algorithmic}
\STATE ALGORITHM:
\STATE $\delta = min((|\vec{N}| + |\overrightarrow{m_3}|) \cdot 0.1, K_n \cdot 0.1)$
\WHILE{$(F, C_1)~\cdot~(F, C_2) <=~0$ \OR $(F, C_1)~\cdot~(F, C_3) <=~0$  \OR $(F, C_1)~\cdot~(F, C_4) <=~0$}
\STATE $K_n = K_n + \delta$
\STATE $F = K_n \cdot \overrightarrow{N} - \overrightarrow{m_3}$
\STATE $\delta = 2 \cdot \delta$
\ENDWHILE
\end{algorithmic}
The proposed method does not pretend to be the most optimal, and in the future it makes sense to study the optimal initial value of $\delta$ and the iterative parameter $K_n$ change function.
	\subsection{Approximation of volumes}
In nonplanar reflection, when constructing a new reflected projection matrix, the volume specified by the 4 frustum reflected lateral planes and 1 reflecting plane cannot be captured exactly by one affine frustum; in this regard, it is proposed to consider an optimization method for approximating this volume.
\\Consider the following problem statement: given 4 side target planes $\overrightarrow{T_0}, \overrightarrow{T_1},\overrightarrow{T_2}, \overrightarrow{T_3} $ are passing in pairs clockwise through the near plane $\vec{N}$ at points $\overrightarrow{X_0}, \overrightarrow{X_1}, \overrightarrow{X_2}, \overrightarrow{X_3}$. It is necessary to find such an optimal frustum origin $O = O(x,y,z)$ that the normals $\vec{n_0}, \vec{n_1}, \vec{n_2}, \vec{n_3}$ of the side planes formed by the corresponding points $\overrightarrow{X_i}, \overrightarrow{X_{i+1}}$ will be closest to the unit normals $\overrightarrow{t_0}, \overrightarrow{t_1},\overrightarrow{t_2}, \overrightarrow{t_3}$ of the target planes.	
$$\vec{n_i} = \overrightarrow{OX_i} \times \overrightarrow{OX_{i+1}},$$
Define the partially functions $F_i = \frac{(\overrightarrow{n_i}, \overrightarrow{t_i})}{|\overrightarrow{n_i|}} = cos(\alpha_i), i = \overline{0, 3},$
\\Optimization problem formulation: $ \max_{(x, y, z)} F = \sum_{i = 0}^{i < 4} F_i^2$\\
\\Objective function gradient:\\
$\nabla F_x = 2 \cdot |\overrightarrow{n_i}|^2 \cdot (\overrightarrow{n_i}, \overrightarrow{t_i}) \cdot (\frac{\partial \overrightarrow{n_i}}{\partial x},\overrightarrow{t_i}) - \frac{\partial |\overrightarrow{n_i}|}{\partial x} \cdot (\overrightarrow{n_i}, \overrightarrow{t_i})^2$
\\
$\nabla F_y = 2 \cdot |\overrightarrow{n_i}|^2 \cdot (\overrightarrow{n_i}, \overrightarrow{t_i}) \cdot (\frac{\partial \overrightarrow{n_i}}{\partial y},\overrightarrow{t_i}) - \frac{\partial |\overrightarrow{n_i}|}{\partial y} \cdot (\overrightarrow{n_i}, \overrightarrow{t_i})^2$
\\
$\nabla F_z = 2 \cdot |\overrightarrow{n_i}|^2 \cdot (\overrightarrow{n_i}, \overrightarrow{t_i}) \cdot (\frac{\partial \overrightarrow{n_i}}{\partial z},\overrightarrow{t_i}) - \frac{\partial |\overrightarrow{n_i}|}{\partial z} \cdot (\overrightarrow{n_i}, \overrightarrow{t_i})^2$
\\
Further, an iterative solution can be obtained using the gradient descent method with the initial value $O(x_0, y_0, z_0)=intersect(\overrightarrow{T_0}, \overrightarrow{T_1}, \overrightarrow{T_2}$).
\section{Affine frustum modifications}
This chapter proposes to consider various operations on frustums that can be used, for example, to efficiently and quickly create reflection and refraction. Also, in order to avoid expensive calculations for inverting the projection matrix of a new transformed frustum, this paper proposes ways to transform the original inverse projection matrix to obtain a new one.
\subsection{ Viewing frustum cutting}
\begin{figure}[ht]
	\begin{center}
		\scalebox{0.3}{
			\includegraphics{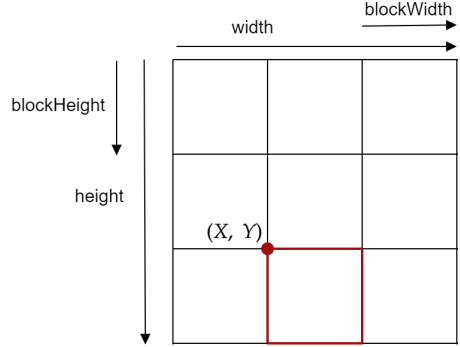}
		}		
		\caption{
			\label{cut}
			Frustum cutting.}
		\end {center}
		\end {figure}

The problem of pruning an affine frustum can be formulated as follows:
in a given frustum $F$ with the dimensions of a rectangular near plane $(width, height)$, it is necessary to obtain the projection matrix of the new frustum $f$ cut from the original by offsetting the upper left corner
by $(X, Y)$ with the new dimensions of the rectangular near plane $(blockWidth, blockHeight)$. The illustration is shown in Figure~\ref{cut}.
To obtain the new projection matrix $P'$ and its inverse one~$P'^{-1}$, it is necessary to produce the following calculations:
$$x = \frac{X}{blockWidth}, y = \frac{Y}{blockHeight}$$

$$ w = \frac{width}{blockWidth}, h = \frac{height}{blockHeight}$$
$$rw = \frac{blockWidth}{width}, rh = \frac{blockHeight}{height}$$
$$
	cropMat = \begin{pmatrix}
		w & 0 & 0 & 0\\
		0 & h & 0 & 0\\
		0 & 0 & 1 & 0\\
		w - 1 - 2x & h - 1 - 2y & 0 & 1
\end{pmatrix}
$$
$$
	invCropMat = \begin{pmatrix}
	rw & 0& 0& 0\\
	0& rh& 0& 0\\
	0& 0& 1& 0\\
	-1 + rw + 2x \cdot rw & -1 + rh + 2y \cdot rh & 0& 1
	\end{pmatrix}
$$
\begin{equation}
	\begin{cases}
	P' = P * cropMat\\
	(P')^{-1} = invCropMat * P^{-1} \label{eq:14}
\end{cases}
\end{equation}
Due to the sparseness of the $cropMat$ and $invCropMat$ matrices, matrix multiplications \eqref{eq:14} can be implemented more efficiently using the following formulas:
\begin{equation}
	\begin{cases}
P' = \begin{pmatrix}w\vec{p_0} +(w - 1 - 2x)\vec{p_3}, & h\vec{p_1} +(h - 1-2y)\vec{p_3}, & \vec{p_2}, & \vec{p_3}\end{pmatrix}\\

(P')^{-1} = \begin{pmatrix}rw\vec{p_0}',& rh\vec{p_1}',& \vec{p_2}', &(-1 + rw + 2x\cdot rw)\vec{p_0}' + (-1 + rh + 2y\cdot rh)\vec{p_1}' + \vec{p_3'}\end{pmatrix}
\end{cases}
\end{equation}
Here $\vec{p_i}$ are the columns of matrix $P$ and $\vec{p_i}'$ are the rows of matrix $P^{-1}$.
\subsection{ Flat viewing frustum reflection}
\begin{figure}[ht]
	\begin{center}
		\scalebox{0.2}{
			\includegraphics{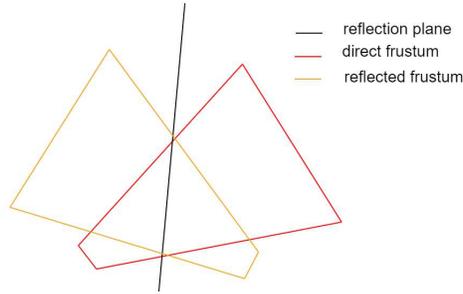}
		}		
		\caption{
			\label{refl}
			Frustum reflection.}
		\end {center}
		\end {figure}

\begin{figure}[ht]
\begin{center}
	\scalebox{0.2}{
		\includegraphics{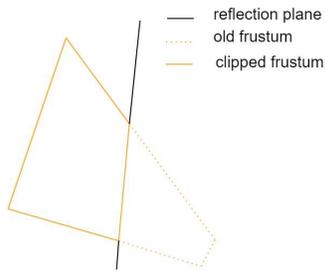}
	}		
	\caption{
		\label{clipping}
		Frustum clipping.}
	\end {center}
	\end {figure}

Reflection on a flat surface consists of two stages:
\begin{enumerate}
	\item Reflection of frustum from a flat surface by multiplying its projection matrix by the reflection matrix (Figure~\ref{refl})
	\item Cutting off frustum by a reflective plane to build a new near plane and subsequent modification of two corresponding frustum matrices (Figure~\ref{clipping})
\end{enumerate}
To reflect frustum from an arbitrary plane $c(c_x, c_y, c_z, c_w)$, $|(c_x c_y c_z)| = 1$ the following matrix is used\cite{fourth}:
\begin{equation}
	R = \begin{pmatrix} 1 - 2c_x^2, & -2c_xc_y, & -2c_xc_z & 0\\ -2c_yc_x, & 1 - 2c_y^2, & -2c_yc_z &0\\ 
		- 2c_zc_x, & -2c_zc_y, & 1-2c_z^2 & 0\\
		- 2c_wc_x, & -2c_wc_y, & -2c_wc_z & 1 \label{eq:16}
	\end{pmatrix}
\end{equation}
\\Thereby, the new perspective projection matrix of reflected frustum $P'$ and its inverse matrix $P'^{-1}$ can be found from the relations:
$$	P' = R * P$$

$$	P'^{-1} = P^{-1} * R$$

The next step is to clip the frustum with a reflective plane so as not to capture the extra volume of the scene. This has already been done in work \cite{second}, however, in our method, we propose a way to modify not only the direct projection matrix $P'$, but also the inverse one $P'^{-1}$.
\\The plane vector $c$ must be such that the camera position $O$ lies in the negative half-plane:
\begin{equation}
	c_xO_x + c_yO_y + c_zO_z + c_w < 0,
\end{equation}
\\Next, we select the point $Q$, farthest from the camera position through which the far plane will pass and construct new matrix projection $P_{new}$:
\begin{equation}
	Q = (sgn(c_x), sgn(c_y), 1., 1.) * (P')^{-1}
\end{equation}
\begin{equation}
	k = \frac{\vec{m_3} \cdot \vec{Q}}{\vec{N} \cdot \vec{Q}}
\end{equation}
\begin{equation}
	P_{new} = \begin{pmatrix}P'[0], & P'[1] & k*\vec{N} & P'[3] \end{pmatrix}, 
\end{equation}
where $P'[i]$ - columns of projection matrix $ P'$.
\\ And the last thing left to do is calculate the inverse projection matrix for $P''$:
\begin{equation}
	 P_{new} = P' * T 
\end{equation}
\begin{equation}
	T = \begin{pmatrix}
		1 & 0 & X_1 & 0 \\ 
		0 & 1 & X_2 & 0 \\ 
		0 & 0 & X_3 & 0 \\
		0 & 0 & X_4 & 1
	\end{pmatrix}
\end{equation}
\begin{equation}
	P'*\vec{X} =  k*\vec{N}
\end{equation}
\begin{equation}
	\vec{X} =  (P')^{-1} * k\vec{N}
\end{equation}
\begin{equation}
	T^{-1} =  \begin{pmatrix}
		1 & 0 & \frac{-X_1}{X_3} & 0 \\ 
		0 & 1 & \frac{-X_2}{X_3} & 0 \\ 
		0 & 0 & \frac{1}{X_3} & 0 \\
		0 & 0 & \frac{-X_4}{X_3} & 1
	\end{pmatrix}
\end{equation}
\begin{equation}
		P_{new}^{-1} =  T^{-1} * (P')^{-1}
\end{equation}
In this way, the final formula for calculating the inverse matrix for the new projection matrix $P_{new}$ has the form:
\begin{equation}
		P_{new}^{-1} = \begin{pmatrix}
			\vec{p_0'} - \frac{X_1}{X_3}\vec{p_2'}\\
			\vec{p_1'} - \frac{X_2}{X_3}\vec{p_2'}\\
			\frac{1}{X_3}\vec{p_2'} \\ 
			\vec{p_3'} - \frac{X_4}{X_3}\vec{p_2'}\\
		\end{pmatrix}
\end{equation}
where $\vec{p_i'}$ are i-th rows of matrix $(P')^{-1}$
\subsection{Flat frustum reflection from rectangle in 3D space}
This chapter proposes to consider a special case of plane reflection: reflection from a rectangular surface. The mathematical description of this problem is as follows: for a given reflective rectangular surface defined by point $P_1$ and two side vectors $\vec{a_1}, \vec{a_2}$ in world space and given coordinates of the position of the observer $O$, construct an affine frustum, which imitates reflection from a given surface in accordance with the point of observation.
\\To construct the projection matrix $M$ of the reflected affine frustum, it is proposed to apply the well-known formula \eqref{eq:28} used in OpenGl. To calculate the unknown parameters n, r, l, t, b, it is necessary to transform the corner points of a rectangle $ P_1, P_2, P_3, P_4 $ \eqref{eq:29} into a new coordinate system associated with the reflected frustum and calculate the edge values along each axes. The missing f value is selected manually with only one condition: $f >n $.
\begin{equation}
 	M = \begin{pmatrix}
 	\frac{2n}{r-l}  & 0 & 0  & 0\\
 	0 & \frac{2n}{t-b} & 0 & 0\\
 	\frac{r+l}{r-l} & \frac{t+b}{t-b} & -\frac{f+n}{f-n} & -1\\
 	0 & 0 & -\frac{2fn}{f-n} & 0  \label{eq:28} 
 \end{pmatrix}
\end{equation}
\begin{equation}
	\begin{cases}
	P_2 = P_1 + \vec{a_1} \\ P_3 = P_2 + \vec{a_2} \\ P_4 = P_1 + \vec{a_2} \label{eq:29} 
	\end{cases}
	\end{equation}
The origin of the new coordinate system is at the point of reflection of the observer $ O'$, which can be calculated by multiplying by the reflection matrix $R$ from formula \eqref{eq:16}, into which the vector of the reflecting plane $\vec{N}$ is substituted \eqref{eq:30}.
\begin{equation}
	\begin{cases}
 \overrightarrow{N_{xyz}} = \frac{\vec{a_1} \times \vec{a_2}}{|\vec{a_1} \times \vec{a_2}|}\\
N_w = -(N_{xyz}, P_1)\\
 O' = R(N) *  O \label{eq:30}
\end{cases}
\end{equation}
According to the rules for constructing the projection matrix, the negative $OZ_-$ axis of the new coordinate system should pass through the middle of the rectangle $C$, and the $Y$-axis is codirectional with the vector $\vec{a_1}$. Thus, the unit vectors $\overrightarrow{i_{new}}, \overrightarrow{j_{new}}, \overrightarrow{k_{new}}$ of the new coordinate system can be obtained by the following formulas:
\begin{equation} C = \frac{P_2 + P_4}{2} 
\end{equation}
\begin{equation}
	\begin{cases}
\overrightarrow{k_{new}} = \frac{\overrightarrow{CO'}}{|\overrightarrow{CO'}|}\\
\overrightarrow{j_{new}} =  \frac{\vec{a_1}}{|\vec{a_1}|}\\
\overrightarrow{i_{new}} = \overrightarrow{j_{new}} \times \overrightarrow{k_{new}}
\end{cases}
\end{equation}
Using the transformation matrix Q, which transforms the coordinates of points from the old coordinate 
system to the new one, which is described in detail in \cite{sixth}, and the shift matrix $T$, the 
required projection matrix $M_{refl}$ is written as follows:
\begin{equation}
		\begin{cases}
			Q = \begin{pmatrix}
i_{old} \cdot i_{new} & j_{old} \cdot i_{new} & k_{old} \cdot i_{new} \\
i_{old} \cdot j_{new} & j_{old} \cdot j_{new} & k_{old} \cdot j_{new} \\
i_{old} \cdot k_{new} & j_{old} \cdot k_{new} & k_{old} \cdot k_{new} 
			\end{pmatrix}\\
			T = \begin{pmatrix}
			1& 0  & 0 &0 \\
			0& 1  & 0 &0 \\
			0& 0  & 1& 0\\
			-O'_x & -O'_y &-O'_z &1
			\end{pmatrix}\\
		M_{refl} = T * Q^{-1} * M\\
		M_{refl}^{-1} = M^{-1} * Q * T^{-1}\\
				\end{cases}
\end{equation}
Since $Q$ is orthogonal then: $Q^{-1} = Q^T$.
\subsection{Simulation of refraction from lens}
\label{section:refraction}
\begin{figure}[ht]
	\begin{center}
		\scalebox{0.25}{
			\includegraphics{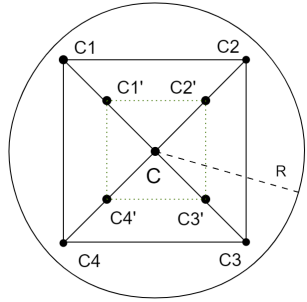}
		}		
		\caption{Uniform grid transformation
			\label{circ}
		}
		\end {center}
		\end {figure}

\begin{figure}[h]
	\centering
	\begin{minipage}{0.4\textwidth}
		\includegraphics[width=\linewidth]{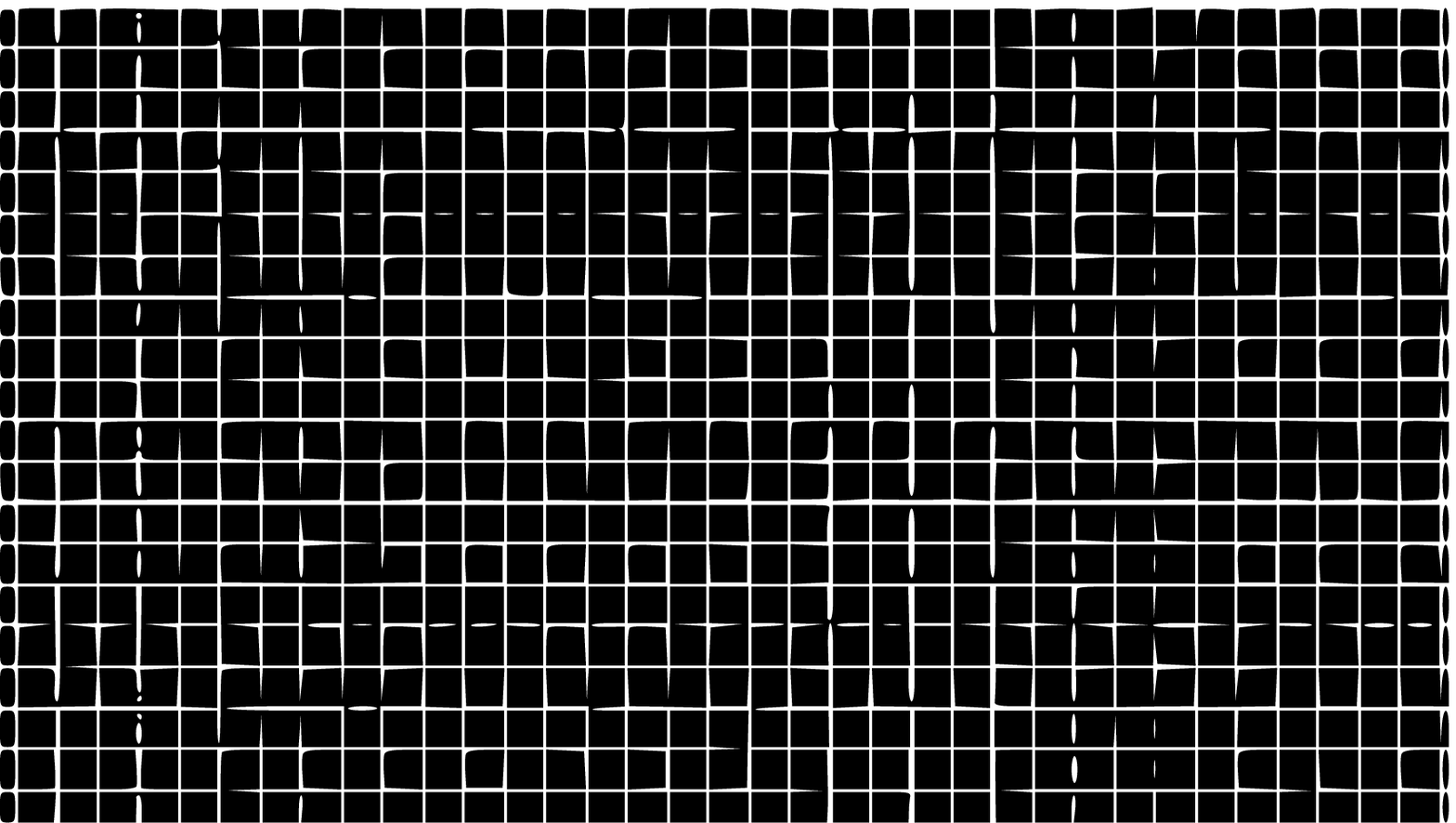}
		\caption{}\label{grid1}
	\end{minipage}
	\begin{minipage}{0.4\textwidth}
		\includegraphics[width=\linewidth]{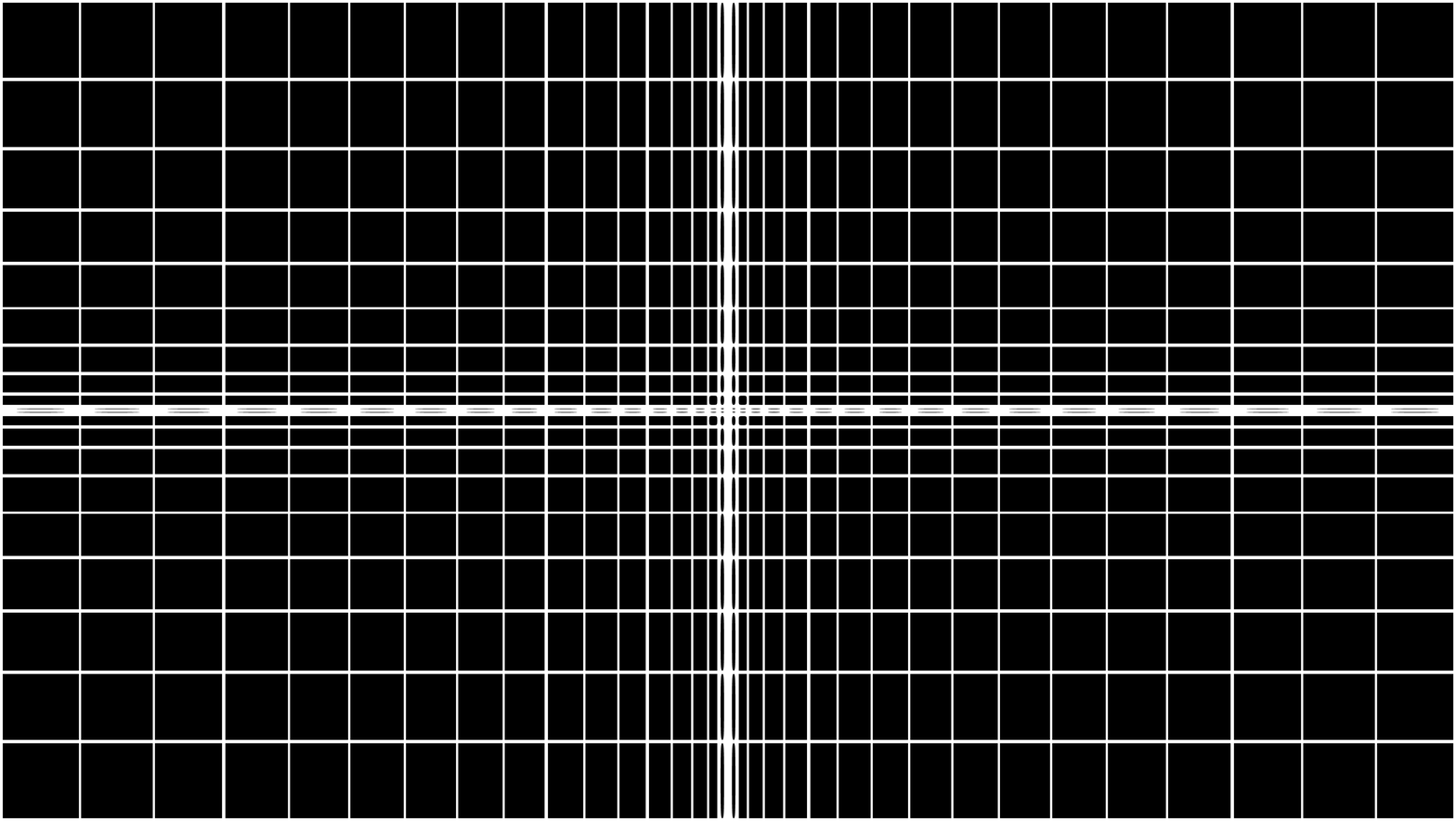}
		\caption{}\label{grid2}
	\end{minipage}
\end{figure}

This section proposes to consider the technique of simulating lenticular refraction in a tile-based rendering system\cite{seventh}. The suggested method consists of the implementation of 2 effects: barrel distortion and magnification (we assume that the lens is collecting).
\\
The key idea of the algorithm is to move 4 corner intersection points (Picture~\ref{circ}) with the refraction plane ($C_1, C_2, C_3, C_4$) to new one ($C_1', C_2', C_3', C_4'$). This transformation \eqref{eq:34} is performed in such way that the points in radius $R$ from the given lens center $C$ approach it while other points outside this radius move away in accordance with the power law with parameter $p$ (degree of the distortion). The result of uniform grid transformation (Picture~\ref{grid1}) is shown in Picture~\ref{grid2}.
		
		\begin{equation}
			\begin{cases}
				\vec{v_i} = \overrightarrow{CC_i}\\
				r_i = \frac{|\vec{v_i}|}{R}\\
				
				\vec{C_i'} = \vec{C} + R \cdot r_i^{p} \cdot \frac{\vec{v_i}}{|\vec{v_i}|}, i = \overline{1, 4}
			\end{cases}\label{eq:34}
		\end{equation}

The next stage is the formation of new frustums as follows: their near plane must coincide with the refraction plane and the side planes form using the new coordinates of the appropriate intersection points. It is also worth noting that all frustums in radius $R$ from $C$ are magnifying because a decrease in the area of the near plane of the constructed frustum automatically entails an increase effect.
\subsection{Frustum validation}
\label{section:validation}
Frustum is considered valid if the near plane is a convex quadrilateral, that separates the origin of the frustum and its body. Convexity testing is a trivial task that can be solved using pairwise vector products of the sides of the polygon. The test for separability can be carried out by substituting the coordinates of the far corner points and the origin of the frustum into the equation of the near plane and checking the signs of distance from points to planes.
\section{Future work}
\label{section:nonAffine}
One of the main inconveniences of using affine frustums is the presence of a relationship between 6 planes, which does not always allow moving the far plane far enough. Also, a strict restriction on the convergence of 4 planes at one point makes it impossible to use affine frustums within non-planar reflection in tile-rendering systems.
To replace the traditional method of translating the point coordinates from three-dimensional space into clip space using the projection matrix, it is proposed to consider the following transformation.
For an arbitrary point $\vec{X}$ in 3D space and given non-affine frustum, represented as a set of 6 planes: left, right, top, bottom, near, far, denoted as $\vec{L}, \vec{R}, \vec{T}, \vec{B}, \vec{N} , \vec{F}$ respectively, the new point coordinates $X_{clip}$ in the clip space can be determined as follows:
\begin{equation}
	X_{clip} = \begin{pmatrix} \frac{d_l}{(d_l + d_r)} & \frac{d_t}{(d_t + d_b)} & \frac{d_n}{(d_n + d_f)}
	\end{pmatrix}\label{eq:36}
\end{equation}
where $d_i$ - signed distance from point $X$ to corresponding plane $i$.
Also it should be noted that planes normals must be oriented outward from the frustum.
\\However, this method is not ideal due to the existence of points that turn the denominator of at least one of the three fractions to zero in~\eqref{eq:36}, which leads to distortions during rasterization and it is necessary to further investigate its possible modifications.
\section{Conclusion}
\label{section:Conclusion}
In this paper, the properties of the projection matrix and the corresponding viewing frustum were studied. In addition, formulas for obtaining frustum key points from the inverse projection matrix were derived. Methods of constructing frustum by given points and planes as well as frustum transformation formulas were considered to simulate such effects as planar and nonplanar reflection, refraction and cropping. In conclusion, the new idea of non-affine frustum application to limit the visible volume and subsequent rasterization were proposed. 
\section{ACKNOWLEDGMENTS}
This project was supported by Huawei Russian Research Institute.	

\end{document}